# Gravitational Collapse and Equilibrium Conditions of a Toroidal Vortex with Thermal Pressure

K.Yu. Bliokh and V.M. Kontorovich

*Institute of Radio Astronomy, 4 Krasnoznamyonnaya st., Kharkov, 61002, Ukraine*
*e-mail: vkont@ira.kharkov.ua*

We found equilibrium conditions for a self-gravitating toroidal vortex by taking into account thermal pressure. These conditions are shown to significantly differ from those for a disk and a sphere. The evolution of a thin vortex turns it into a compact vortex that loses mechanical stability for low masses at a polytropic index $\gamma < 4/3$ but retains stability for sufficiently high masses and densities determined by the velocity circulation in the vortex.

## 1. Introduction

Obviously, the physical stability conditions for a self-gravitating vortex torus (Fig. 1) fundamentally differ from those for a disk. First of all, the constancy of the circulation $\Gamma = 2\pi r v$ in the vortex results in the maximum velocity at the smallest (inner) radius of the torus, and the condition imposed by it on the angular velocity $\Omega(r) = \Gamma / 2\pi r^2$ results in the absence of epicyclic frequency $[\nu(r)]^2 = \frac{1}{r^3}\frac{d}{dr}\left[r^2\Omega(r)\right]^2$ in the main approximation. Whereas during the rotation of a disk or a sphere, the detaching particles escape from the equator (to form an additional disk in the case of a sphere), the particles in a toroidal vortex detach the minimum radius and escape to form a one-sided jet[1] even in the absence of a magnetic fields (Bliokh and Kontorovich 2003; Shatskii and Kardashev 2002; Ansorg *et al.* 2002). In addition to the detachment due to particle collisions in the vortex torus, which is facilitated by the self-intersecting trajectories in the effective potential of the vortex (Bliokh and Kontorovich2003), the particles can be effectively accelerated along the torus axis by the induction field in the case of a magnetized torus (Shatskii and Kardashev 2002).

The discovery of obscuring tori (Antonucci 1993) and their direct observation both in galactic (Risaliti *et al.* 2003) and stellar objects (see Bogovalov and Khangulyan (2002) for references) makes the problem of studying gravitating toroidal vortices particularly relevant. As far as we know, until recently, this problem has not been considered (for references, see Bliokh and Kontorovich (2003), who made the first attempt). In this paper, we extend our results (Bliokh and Kontorovich2003) to finite temperatures and take into account the effects of thermal pressure. As we will see below, this will affect the equilibrium and collapse conditions.

## 2. Equations of motion

Previously (Bliokh and Kontorovich 2003), based on the Hamiltonian dynamics, we derived the equations that described the evolution of a thin selfgravitating toroidal vortex:

---

[1] The excitation of a unidirectional flow by a magnetized rotating torus was considered by Bisnovatyi-Kogan (1989) as a possible formation mechanism of cosmic jets (Ustyugova *et al.* 2000). For a qualitative association of one-sided jets changing direction with rotating binary galactic nuclei, see Zhuravlev and Komberg (1999).

$$\ddot{r} = \frac{p_\varphi^2}{M^2 r^3} - G\frac{M}{\pi R r} \; ,$$

$$\ddot{R} = -G\frac{M}{\pi R^2}\ln\frac{\alpha r}{R} \; .$$

(1)

Here, $r$ and $R$ are the small and large radii of the torus, $M$ is the total vortex mass, $G$ is the gravitational constant, $\alpha \sim 1$ is a numeric factor, and $p_\varphi = Mr^2\dot\varphi = const$ is the angular momentum of the matter ($\varphi$ is the cyclic rotation coordinate at the small radius).

To introduce the thermal pressure forces into the equations of motion (1), we note that the virtual work done by the external forces against the pressure force is

$$\delta A = -\delta E = p\delta V \; ,$$

(2)

where $p$ is the pressure in the vortex, and $V$ is the vortex volume (below, we disregard the distribution of matter and pressure at the small torus radius while ignoring the factors of the order of unity in the formulas). Since the torus volume is $V = 2\pi^2 r^2 R$, we obtain from (2)

$$-\delta E = 4\pi^2 r R p \delta r + 2\pi^2 r^2 p \delta R \; .$$

(3)

Hence, we find the forces acting along the $r$ and $R$ coordinates $F_r = -\delta E/\delta r$ and $F_R = -\delta E/\delta R$ and, accordingly, the additional terms in the pressure forces in (1):

$$\ddot{r} = \frac{p_\varphi^2}{M^2 r^3} - G\frac{M}{\pi R r} + \frac{p}{M}4\pi^2 R r \; ,$$

$$\ddot{R} = -G\frac{M}{\pi R^2}\ln\frac{\alpha r}{R} + \frac{p}{M}2\pi^2 r^2 \; .$$

(4)

To relate the pressure to the vortex parameters, we must invoke the equation of state for the matter and the vortex evolution. We assume that the gas is perfect and that the compression (expansion) is adiabatic. Introducing the initial temperature of the matter $T_0$, the initial vortex parameters $r_0$ and $R_0$, and the mass of one gas particle $m_0$, we then obtain

$$p = \frac{p_0 V_0^\gamma}{V^\gamma} = \frac{M T_0}{m_0}\frac{(2\pi^2 r_0^2 R_0)^{\gamma-1}}{(2\pi^2 r^2 R)^\gamma} \; .$$

Substituting this expression into (4), we obtain instead of (1)

$$\ddot{r} = \frac{p_\varphi^2}{M^2 r^3} - G\frac{M}{\pi R r} + \frac{T_0}{m_0}\frac{2(r_0^2 R_0)^{\gamma-1}}{r^{2\gamma-1} R^{\gamma-1}} \; ,$$

$$\ddot{R} = -G\frac{M}{\pi R^2}\ln\frac{\alpha r}{R} + \frac{T_0}{m_0}\frac{(r_0^2 R_0)^{\gamma-1}}{r^{2\gamma-2} R^\gamma} \; .$$

(5)

## 3. Equilibrium conditions

As previously (Bliokh and Kontorovich 2003), we assume that for a thin torus $r \ll R$, the evolution (Fig. 2) at the small radius is much faster than it is at the large radius, and that a quasi-equilibrium state in $r$ is initially established. This state is determined by the fact that the force on the right-hand side of the first equation (5) vanishes:

$$\frac{p_\varphi^2}{M^2 r^3} - G\frac{M}{\pi R r} + \frac{T_0}{m_0}\frac{2(r_0^2 R_0)^{\gamma-1}}{r^{2\gamma-1} R^{\gamma-1}} = 0 \; .$$

(6)

This equation is not solvable directly and implicitly specifies the relation $r(R)$. However, if we introduce the linear mass per unit length of the torus ring $\chi = M/2\pi R$, the mean density $\rho = M / 2\pi^2 r^2 R$, and the velocity circulation $\Gamma = 2\pi p_\varphi / M$ and consider (6) as the equation for the linear Jeans mass $\chi(\rho)$, then an explicit expression follows from it:

$$\chi(\rho) = \frac{T_0}{2Gm_0} \frac{\rho^{\gamma-1}}{\rho_0^{\gamma-1}} \left\{ 1 + \sqrt{1 + \xi \rho^{3-2\gamma} \Gamma^2} \right\} \text{, where } \xi = \frac{G}{2\pi} \left( \frac{m_0 \rho_0^{\gamma-1}}{T_0} \right)^2. \tag{7}$$

Since $\chi = \pi r^2 \rho$, it is convenient to consider formula (7) in the same way as the formula for the small vortex radius in the equilibrium conditions at the first evolutionary stage.

Equating $\chi(\rho) = \chi$ and inverting relation (7) yields an equation for the equilibrium density at the specified linear mass $\chi$ and velocity circulation $\Gamma$. Without analyzing $\chi(\rho)$ for reasons of space, we note that an equilibrium solution always exists (Fig. 3) if the density does not exceed $\rho_*$:

$$\rho < \rho_* = \left( f_* / 2 \right)^{\frac{1}{\gamma-1}}, \tag{8'}$$

where $f_* = \nu \chi$ is proportional to the linear mass of the vortex, $\nu = 2G \left( m_0 \rho_0^{\gamma-1} / T_0 \right)$.

In this case, the equilibrium density satisfies the transcendental equation

$$\xi \Gamma^2 = \frac{f_* \left( f_* - 2\rho^{\gamma-1} \right)}{\rho}, \tag{8}$$

whose right-hand side degenerates into a hyperbola for $\gamma = 1$ or into a hyperbola shifted in the vertical axis for $\gamma = 2$.

Once the local equilibrium (6) has been established, the vortex will slowly evolve at the large radius $R$ with the conservation of the relation $r(R)$. If the force on the right-hand side of the second equation (5) is assumed to be negative, then the contraction will continue until this force vanishes:

$$-G \frac{M}{\pi R^2} \ln \frac{\alpha r}{R} + \frac{T_0}{m_0} \frac{\left( r_0^2 R_0 \right)^{\gamma-1}}{r^{2\gamma-2} R^\gamma} = 0. \tag{9}$$

If the thermal pressure forces are small compared to the centrifugal forces, then the first two terms in (6) compete, and the thermal pressure is also low compared to the gravitational forces. In this case, we obtain the square-root dependence $r \sim \sqrt{R}$, considered previously (Bliokh and Kontorovich 2003). The contraction at the large radius gradually overtakes the contraction at the small radius, and there comes a time when they become of the same order of magnitude, $r \sim R$. This order-of-magnitude relation also corresponds to equilibrium (9): in this region, the gravitational force is balanced by an arbitrarily weak thermal pressure, because the logarithm is small. If, however, the thermal pressure forces are large compared to the centrifugal forces, then the last two terms in (6) compete. As can be easily seen, their balance corresponds to the equilibrium condition (9), where $\ln(\alpha r / R) \sim 1$. Obviously, this case also corresponds to $r \sim R$.

Thus, irrespective of the thermal pressure, the vortex equilibrium is established in the region where both torus radii become of the same order of magnitude: $r \sim R$. Using dependence (6), we obtain the following equation for the equilibrium parameters:

$$\frac{p_\varphi^2}{M^2 R^3} - G \frac{M}{\pi R^2} + \frac{T_0}{m_0} \frac{2 \left( r_0^2 R_0 \right)^{\gamma-1}}{R^{3\gamma-2}} = 0. \tag{10}$$

This equation defines the characteristic radius $R$ of a compact vortex that came to an equilibrium: the balance between the compressing gravitational forces and the extending centrifugal and thermal pressure forces[2].

---

[2] Actually, in the region $r \sim R$, the initial approximation of a thin ring torus and the separation of evolution scales in $r$ and $R$ becomes invalid. Here, we must consider a single compact object that resembles the Hill vortex rather than the Maxwell vortex. This can be easily done by analogy with our previous study (Bliokha nd Kontorovich 2003) by including the pressure forces for a

Equation (10), like (6), is not explicitly solvable for $R$. We may consider two limiting cases. When the pressure forces are small (the first term is much larger than the third term), we come to the equilibrium between the gravitational and centrifugal forces considered previously (Bliokh and Kontorovich 2003). In this case,

$$R \sim \frac{p_\varphi^2}{GM^3} \:. \tag{11}$$

In the opposite limit, when the centrifugal forces are small (the first term in (10) is much smaller than the third term), we come to the equilibrium between the gravitational and pressure forces:

$$R^{3\gamma-4} \sim \frac{T_0 \left(r_0^2 R_0\right)^{\gamma-1}}{GMm_0} \sim \frac{T_0 V_0^{\gamma-1}}{GMm_0} \:. \tag{12}$$

The limiting value of (11) is obtained for the parameters that satisfy

$$\frac{T_0 V_0^{\gamma-1}}{m_0} \frac{G^{3\gamma-5} M^{9\gamma-13}}{p_\varphi^{6\gamma-8}} \ll 1 \:, \tag{13}$$

while the limiting value of (12) is obtained for the inverse inequality.

## 4. Analysis of the equilibrium conditions: equilibrium and non-equilibrium Jeans masses

In the equilibrium equation (10), we change from the radius $R$ and the angular momentum $p_\varphi$ to the density $\rho \sim M/R^3$ and the circulation $\Gamma = p_\varphi / M$. Equation (10) then transforms into an equation for the generalized Jeans mass $M(\rho)$:

$$GM^{4/3} - \frac{T_0}{m_0} \frac{\rho^{\gamma-4/3}}{\rho_0^{\gamma-1}} M^{2/3} - \Gamma^2 \rho^{1/3} = 0 \:, \tag{14}$$

where we omitted the dimensionless factors of the order of unity, which will be restored below. Solving the equation yields $M^{2/3}(\rho)$:

$$\nu M^{2/3}(\rho) = \rho^{\gamma-4/3} \left[ 1 + \sqrt{1 + \xi \rho^{3-2\gamma} \Gamma^2} \right] \:, \text{ where } \nu = \frac{2Gm_0 \rho_0^{\gamma-1}}{T_0} \:,\: \xi = 4G \left( \frac{m_0 \rho_0^{\gamma-1}}{T_0} \right)^2 \:.$$

The equilibrium states are defined by the equality $M = M(\rho)$, where $M$ is the vortex mass. Let us analyze the function $M^{2/3}(\rho)$ and ascertain whether stable equilibrium states can exist.

The equation $dM^{2/3}(\rho)/d\rho = 0$ with $\gamma < 4/3$ gives only one extremum (minimum) of the function $M(\rho)$:

$$\rho_c^{3-2\gamma} = 36\left(\gamma - \frac{4}{3}\right)\left(\gamma - \frac{5}{3}\right) \Big/ \xi \Gamma^2 \:. \tag{15}$$

For $\gamma > 5/3$, there is also a positive root, but it is redundant, as can be seen from the derivation of formula (15) (see Appendix A). The value of $M_c^{2/3} \equiv M^{2/3}(\rho_c)$ for $\gamma < 4/3$ is

$$M_c^{2/3} = \frac{T_0}{2Gm_0} \frac{\rho_c^{\gamma-4/3}}{\rho_0^{\gamma-1}} \left[ 1 + \sqrt{1 + 36\left(\gamma - \frac{4}{3}\right)\left(\gamma - \frac{5}{3}\right)} \right] \:. \tag{16}$$

Note that for $\gamma \to 4/3$ $\rho_c \to 0$ and to calculate $M_c^{2/3}$, we must evaluate the indeterminate form $\lim_{x \to 0+} x^{-x} = 1$. As a result, for $\gamma \to 4/3$, $M_c^{2/3} \to T_0 / Gm_0 \rho_0^{\gamma-1}$. As we see, whereas for $\gamma > 4/3$, the function $M^{2/3}(\rho)$ monotonically decreases with increasing density and there is a

---

spheroidal vortex, as above. However, in this case, we obtain an equation that differs from (6) only by factors of the order of unity, which are anyway unimportant in our rough analysis. Therefore, we omit here these calculations.

solution that describes an equilibrium vortex for any mass, for $\gamma \leq 4/3$, a gap in masses emerges. The equilibrium is possible only for masses larger than some critical mass (16). For $\gamma < 4/3$, this critical mass depends on the circulation $\Gamma$.

Here, we see a significant difference from the gravitational contraction of a nonrotating body, for which no stable equilibrium states are known to exist for $\gamma < 4/3$.

For vortex masses higher than the critical mass $M_c$ that corresponds to the minimum of $M(\rho)$, there are two solutions (Fig. 4). One of them is unstable at densities lower than $\rho_c$ and corresponds to the intersection of the descending branch of the function $M(\rho)$ by the level of a given mass $M$. It is similar to the collapse of a nonrotating mass: the Jeans mass decreases with increasing density, causing the contraction to continue.

In contrast, at densities higher than $\rho_c$, a stable state is realized: the Jeans mass increases as it contracts, and no collapse takes place. This equilibrium state is essentially determined precisely by the rotation of matter in the vortex: the asymptotics of $M^{2/3}(\rho)$ at high densities is $M^{2/3}(\rho) \propto \Gamma \rho^{1/6}$ ($\rho \to \infty$).

## 5. Obtaining the results from evaluation considerations and other approaches

Here, as in our previous paper (Bliokh and Kontorovich 2003), we made good use of the approach of classical Hamiltonian mechanics to solve the problem of the evolution of a toroidal vortex. In this section, we show that the results obtained from the mechanical model can also be obtained from dimension and similarity considerations. In addition, in Appendix B, we show that the hydrodynamic allowance for the thermal pressure yields the same results.

Let us first return to a cold system and consider the final evolutionary stage, when the small and large vortex radii become equal in order of magnitude. In this case, the number of defining parameters decreases to an extent that the characteristic radius $R$ can be determined from the dimension considerations alone. Indeed, such parameters are the conserved mass $M$ and circulation $\Gamma$ (or angular momentum $p_\varphi = M\Gamma$) as well as the gravitational constant $G$ that determines the interaction mechanism (self-gravitation). From these quantities, we can construct only one quantity with the dimensions of length, which can be naturally identified with the Jeans length

$$\lambda = \frac{\Gamma^2}{GM} \ . \qquad (17)$$

Formula (17) is identical to the expression for the equilibrium radius of a compact cold vortex (Bliokh and Kontorovich (2003), formula (23)) if $\Gamma$ means the circulation divided by $2\pi$.

Let us now turn to the early evolutionary stage of a cold vortex, when it is a thin ($r << R$) torus with the local quasi-equilibrium dependence $r = r(R)$. The latter can be derived from the balance between the gravitational $GM/R$ and centrifugal $p_\varphi^2/2Mr^2$ energies:

$$r^2 \sim \frac{\Gamma^2}{GM} R = \lambda R \ . \qquad (18)$$

Thus, $\widetilde{\lambda} = \sqrt{\lambda R}$ acts as the Jeans scale for the small radius. Formula (18) is again identical, to within a factor of about unity, to that derived previously (Bliokh and Kontorovich (2003), formula (14)). The square-root dependence $r \sim \sqrt{\lambda R}$ is related to the corresponding coordinate dependences: $\propto 1/R$ for gravitation and $\propto 1/r^2$ for centrifugal energy. As $R$ decreases, $r \to R$, and we observe the transition to the final evolutionary stage, $\tilde{\lambda} \to \lambda$, in accordance with what was said above[3].

Let us now take into account the thermal motion in the problem under consideration. Restricting our analysis to the late evolutionary stage, when only one scale with the dimensions of length is available, we change in (17)

$$v^2 \to v^2 + v_T^2 \ . \tag{19}$$

where $v$ is the regular velocity, and $v_T$ is the thermal velocity. Since $\Gamma^2 = R^2 v^2$, the following change will occur in $\lambda$:

$$\lambda \to \frac{\Gamma^2 + R^2 v_T^2}{GM} \ . \tag{20}$$

The equilibrium equation $R = \lambda$ now transforms into an equation for $R$ that, after substituting the equations of state for an ideal gas $v_T^2 \propto T$ and a polytropic process $T \propto \rho^{\gamma-1}$ into (20), is identical (to within factors of about unity) to Eq. (10). We also see that for given circulation and temperature, there is a lower limit on the equilibrium mass:

$$M_J \geq \frac{2\Gamma v_T}{G} \ . \tag{21}$$

The equilibrium condition for a toroidal vortex is derived from hydrodynamic equations in Appendix B. In our derivation, we use an analogy with the Jeans dispersion relation for heavy sound:

$$\omega^2 = (k v_T)^2 - \omega_J^2 \ , \quad \omega_J^2 = 4\pi G \rho_0 \ , \tag{22}$$

by extending it to the case of is a regular motion at velocity $v$:

$$\omega^2 = k^2 \left( v_T^2 + v^2 \right) - \omega_J^2 \ . \tag{23}$$

Since suchgeneraliz ations have long been used for a disk (Pikel'ner 1976; Saslaw 1989; Tassoul 1982), we immediately note the difference. The disk geometry changes $\omega_J$ by a factor of $\sqrt{2}$; in addition, the dispersion relation for a disk contains an epicyclic frequency. In our case, this frequency is absent, as was noted in the Introduction. Note also that the dispersion relation (23) contains no term linear in $v$ (responsible for the Doppler effect), because the radial oscillations in the torus tube under consideration (see Appendix B) and the regular velocity are orthogonal to each other.

## 6. Conclusions

In the case under consideration, there are no epicyclic oscillations, because the velocity circulation is constant, which, together with the coordinate independence of the density, significantly simplifies the problem and makes possible its closed analytical study presented here. The possible appearance of jets that carry away the angular momentum must result in the continuation of contraction and collapse even in situations that are stable from the viewpoint of our analysis.

The relativistic generalizations of the problem under discussion, as well as our study, can be useful in astrophysical applications (from young stellar objects and supernovae to active galactic nuclei), in which self-gravitation and rotation are important. It may also be important to take into account the orbital rotation of the torus and the presence of a central mass, whose influence was ignored in this paper (see Bliokh and Kontorovich (2003) for remarks).

---

[3] Note that the inner radius, the radius of the hole of a compact vortex torus that is the difference between the quantities of the same order of $R$, cannot determined by such a rough analysis. Meanwhile, its knowledge is of crucial importance in solving the problem of jet formation and torus collapse.

## Acknowledgment

This work was supported in part by INTAS (grant no. 00-00292).

## Appendix A: Finding the extremum of the function $M(\rho)$

Let us determine the derivative $dM^{2/3}(\rho)/d\rho$ from (14) and set it equal to zero:

$$\frac{dM^{2/3}(\rho)}{d\rho} = \left(\gamma - \frac{4}{3}\right)\rho^{\gamma-\frac{4}{3}-1}\left(1 + \sqrt{1+\xi\Gamma^2\rho^{3-2\gamma}}\right) + \rho^{\gamma-\frac{4}{3}}\frac{\xi\Gamma^2(3-2\gamma)\rho^{3-2\gamma-1}}{2\sqrt{1+\xi\Gamma^2\rho^{3-2\gamma}}} = 0 .$$

Canceling by $\rho^{\gamma-\frac{4}{3}-1}$ and collecting the terms, we obtain

$$-2\left(\gamma - \frac{4}{3}\right)\sqrt{1+\xi\Gamma^2\rho^{3-2\gamma}} = 2\left(\gamma - \frac{4}{3}\right) + \frac{1}{3}\xi\Gamma^2\rho^{3-2\gamma} .$$

We see that for $\gamma > 4/3$, the equality is impossible, because the signs of the right- and left-hand sides differ. Let us introduce $z \equiv \rho^{3-2\gamma}$. For $\gamma < 4/3$, the equality is possible if $2\left(\gamma - \frac{4}{3}\right) + \frac{1}{3}\xi\Gamma^2 z > 0$. After squaring, we find the solution $z_c = \frac{36(\gamma-4/3)(\gamma-5/3)}{\xi\Gamma^2}$, which we can use for $4/3 > \gamma > 4/3 - \xi\Gamma^2 z/2$, according to the above conclusion. As can be easily verified, the derived value of $z_c$ satisfies the required inequality.

## Appendix B: A hydrodynamic approach to the problem

Let us first derive the dispersion relation for acoustic waves in a thin self-gravitating gaseous ring by replacing it by a cylinder with periodic boundary conditions at its ends. The basic equations are the hydrodynamic equations in a self-consistent gravitational field:

$$\frac{\partial \mathbf{v}}{\partial t} + (\mathbf{v}\nabla)\mathbf{v} = \frac{1}{\rho}\nabla p - \nabla\phi , \quad \frac{\partial \rho}{\partial t} + \text{div}\,\rho\mathbf{v} = 0 , \quad \Delta\phi = 4\pi G\rho . \quad (B1)$$

We assume the unperturbed density $\rho_0$ to be independent of the coordinates; i.e. we ignore the effect of the unperturbed potential $\phi_0$, which seems quite justified for a thin torus. For purely radial modes, we then obtain the equation

$$\frac{\partial}{\partial r}\frac{1}{r}\frac{\partial}{\partial r}(r\tilde{v}_r) + \frac{\omega^2 + \omega_J^2}{c_s^2}\tilde{v}_r = 0 . \quad (B2)$$

Here, $\tilde{v}_r$ is the perturbed radial velocity, $\omega$ is the oscillation frequency, $\omega_J = \sqrt{4\pi G\rho_0}$ is the Jeans frequency, and $c_s$ is the speed of sound. The substitution $\frac{\sqrt{\omega^2+\omega_J^2}}{c_s}r \to z$ reduces Eq. (B2) to Bessel equation,

$$z^2\frac{\partial^2 v}{\partial z^2} + z\frac{\partial v}{\partial z} + (z^2 - 1)v = 0 ,$$

for $v(z) = \tilde{v}_r(r(z))$. The solution $J_1(z)$ corresponds to the zero boundary conditions on the cylinder if the following condition is satisfied:

$$\frac{\sqrt{\omega^2+\omega_J^2}}{c_s}r_0 = j_1 ,$$

where $j_1$ is the first zero of the function $J_1(z)$, and $r_0$ is the cylinder radius. Thus, the dispersion relation is

$$\omega^2 = \frac{c_s^2}{r_0^2} j_1^2 - \omega_J^2 \ . \tag{B3}$$

We have omitted the mode index for the frequency: obviously, we can also choose another oscillation mode. However, because of the inequalities for the eigenvalues of the Sturm–Liouville problem, only the contribution of the fundamental nodeless mode is important in studying the stability.

The dispersion relation (B3) includes the finite temperature, but without the vortex motion of matter in the torus. In contrast, the equilibrium conditions at the small radius in a thin vortex obtained previously (Bliokh and Kontorovich (2003)):

$$r = \sqrt{\frac{\pi p_\varphi^2 R}{GM^3}} \ , \tag{B4}$$

include only the regular vortex motion. To take into account the two factors, we note that condition (B4) can be rewritten as

$$\frac{2\Gamma^2}{\pi r^4} - \omega_J^2 = 0 \ . \tag{B5}$$

Thus, comparing (B5) and (B6), we can easily write the equilibrium condition $\omega^2 = 0$ with thermal pressure in the general case of interest to us:

$$\frac{2\Gamma^2}{\pi r^4} + \frac{c_s^2}{r^2} j_1^2 - \omega_J^2 = 0 \ . \tag{B6}$$

The corresponding "dispersion" relation contains $\omega^2$ instead of zero on the right-hand side. Therefore, the negativity of the left-hand side of (B6) corresponds to the instability ($\omega^2 < 0$).

If we express $\omega_J^2$ in terms of the vortex mass and radii, then, after several simple transformations, we can derive an expression for the equilibrium small radius of the torus:

$$r^2 = \frac{4\Gamma^2/\pi^3}{\dfrac{GM}{\pi R} - 2c_s^2 j_1^2} \ . \tag{B7}$$

Since $c_s = c_s(T)$, for isothermal compression, (B7) is the final expression for the quasi-equilibrium small radius that acts as the Jeans scale. We see that an allowance for the finite temperature causes an increase in the Jeans scale (which tends to $\infty$ as $2\pi c_s^2 j_1^2 \to GM/R$) and, accordingly, a deviation from the square-root dependence $r \propto \sqrt{R}$ that arises at $T=0$. Since, in general, $c_s$ indirectly depends on $r$, (B7) is not the solution for $r(R)$, but must itself be considered as an equation. Multiplying (B7) by $2\pi^2 R\rho$, we obtain an equation for the Jeans mass $\chi(\rho)$ per unit length of a thin torus $\chi \equiv M/2\pi R$ that describes the first evolutionary stage of the vortex; for its solution, see in (7) at $\nu = \dfrac{2G}{\zeta\sigma\tau}$, $\xi = \dfrac{32G}{(\pi\zeta\sigma\tau)^2}$ where $\sigma = \dfrac{c_s^2}{T}$ and $\tau = \dfrac{T}{\rho^{\gamma-1}}$.

To within factors of about unity, formulas (B6) and (B7) are identical to the local equilibrium condition (6) when the equation of state for an ideal gas $c_s^2 \propto T$ and a polytropic process $T \propto \rho^{\gamma-1}$ are substituted into them. If we set $r = \varepsilon R$ ($\varepsilon < 1$), then (B6) will naturally lead us to the equilibrium condition (10) and solution (14) for the Jeans mass of a compact vortex

$$\mu = (2/\pi)^{1/3} G/\kappa \ , \quad \xi = 8G/(\kappa\varepsilon\pi)^2 \ , \quad \kappa = \zeta c_s^2 \rho^{1-\gamma} \ . \tag{B8}$$

Thus, both the mechanical and hydrodynamic approaches to the problem under consideration yield identical (within the accuracy under discussion) equations and results.

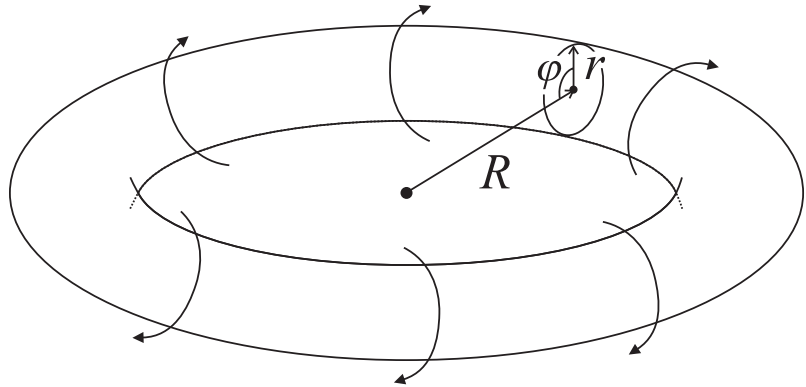

**Fig. 1.** A thin toroidal vortex. The arrows indicate the directions of the streamlines that produce the velocity circulation.

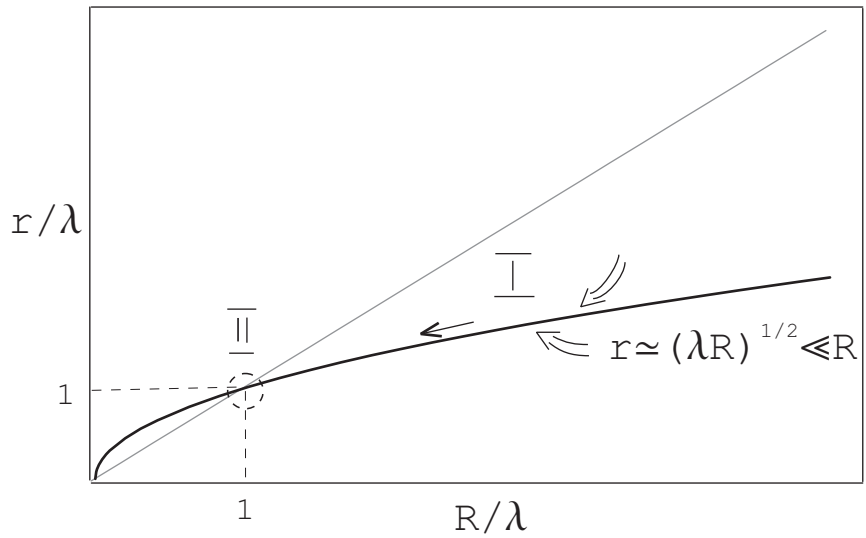

**Fig. 2.** Schematic evolution of a vortex without thermal pressure (Bliokh and Kontorovich 2003).

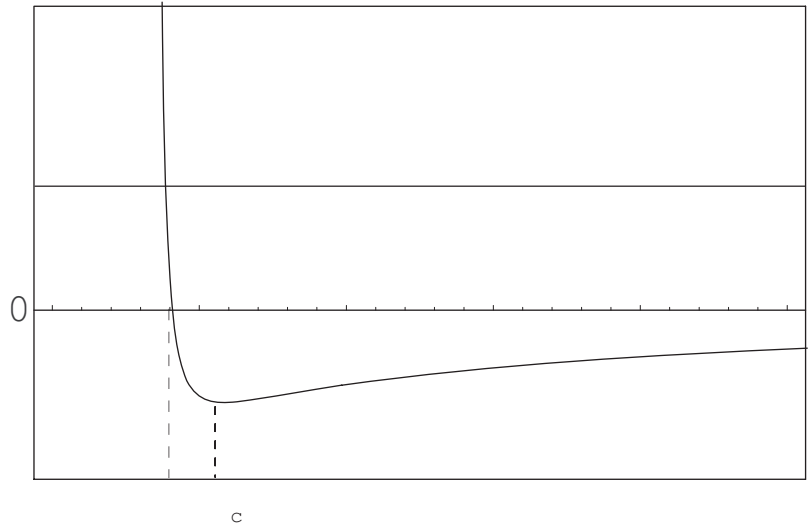

**Fig. 3.** The right-hand side of Eq. (8) for a thin toroidal vortex versus its density.

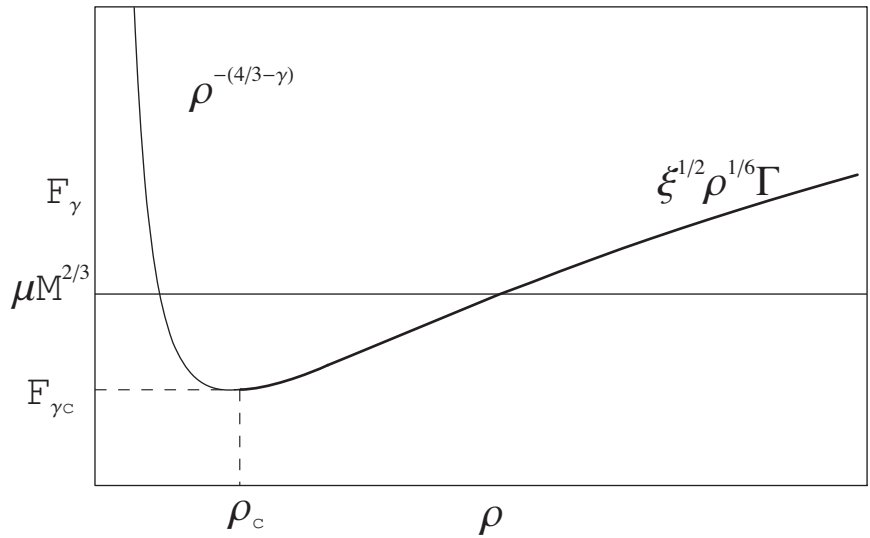

**Fig. 4.** Critical Jeans mass versus torus density for a polytropic index $\gamma < 4/3$. The ascending branch $F_\gamma$ ($\rho > \rho_c$) is stable and is determined by the circulation $\Gamma$.